# Sculpting light with paper for computational mobile phone microscopy

**ANTONY ORTH**[1,2]

[1] *Current address: Computer Vision and Graphics Group, National Research Council of Canada, Ottawa, Ontario, Canada*
[2] *Worked performed while at: The ARC Centre of Excellence for Nanoscale BioPhotonics, School of Science, RMIT University, Melbourne, Victoria, Australia*

*\*email: antony.orth@nrc-cnrc.gc.ca*

**Many optical microscopy techniques rely on structured illumination by way of a projected image (eg. structured illumination microscopy) or a tailored angular distribution of light (eg. ptychography). Electro-optical equipment such as spatial light modulators and light emitting diode (LED) arrays are commonly used to sculpt light for these imaging schemes. However, these high-tech devices are not a requirement for light crafting. Patterned diffusely reflecting surfaces also imprint a spatio-angular structure onto reflected light. We demonstrate that paper patterned with a standard commercial printer can serve to structure the illumination light field for refocusing and stereo microscopy on a mobile phone microscope. Our results illustrate the utility of paper as a light sculpting element for low-tech computational imaging.**

Compact and low-cost computational imaging devices are revolutionizing point-of-care diagnostics, screening, and pathogen detection. On the detection side, these devices often leverage the integrated CMOS sensor and lens in mobile phone camera modules. The mechanical chassis that holds and manipulates the sample can be 3D printed to suit to the intended application. Illumination can be provided in many ways from light emitting diode (LED) arrays for angle-dependent brightfield illumination[1], or side-firing LEDs to couple totally internally reflected light into a microscope slide for fluorescence detection[2]. The internal mobile phone flash and ambient illumination can also be used for brightfield and darkfield microscopy, respectively [3]. Active and electronically addressable external light sources such as LED arrays are a core part of computational imaging approaches such as computational refocusing[4] and Fourier ptychography[5–7], yet add incongruent complexity to mobile imaging platforms. In this Letter we demonstrate that a class of computational imaging techniques relying on angle-structured light can be performed without addressable light source arrays. Instead, we show that an inkjet-printed pattern on paper can act as an effective light structuring element, essentially replacing an LED array placed in the back focal plane of a condenser lens. Though the paper does not "condense" light, it does structure the angular distribution of light passing through a point at a distant sample plane. Colored inks can also be used to further alter the spectral structure of light. Here, we demonstrate both angular and spectral structuring of illumination light fields using paper, for computational refocusing and stereo microscopy.

Previous work has demonstrated that computational refocusing of a weakly absorbing sample can be achieved by controlling the angle of the illumination[4]. In this earlier paper, the sample is illuminated by a computer-controlled LED array, which is placed at a large distance from the sample. Light from each LED in the array impinges on the sample at a particular angle of incidence, as set by the off-axis location of the LED and the distance from the sample to the LED array. We show that this same approach can also be achieved by structuring the illumination using a patterned piece of paper. In order to emphasize the simplicity of the technique, we apply it to mobile phone microscopy, where minimizing bulk, equipment cost, and complexity is of high importance.

Our mobile phone microscope consists of a 3D printed clip-on housing containing a reversed mobile phone lens that acts as the objective lens[8]. The 3D printed clip design, fabrication, and assembly is described in detail in a previous publication[3]. For this work, the 3D printed clip is printed with a Form 2 printer using standard black resin (Formlabs). The 3D printed clip in this Letter differs from previous work by the addition of a thin slot into the back of the clip (Fig. 1a). A piece of paper can be inserted in to the paper slot to pattern light as described below.

The sample is illuminated using the integrated mobile phone flash, via a diffuse reflection off a piece of paper residing in the paper slot. This diffuse reflection occurs D=20mm behind the sample (denoted by P1 and P2 in Fig. 1). Light travels from the flash to the printer paper in a R=5mm diameter tunnel. The paper is white but is made to be nearly all black using an inkjet printer, except for a thin white horizontal stripe of width h=1mm. Light incident on the black part of the paper strip is largely absorbed, while light incident on the white stripe is diffusely reflected towards the sample. The illumination area on the paper strip is a 5mm diameter circle, which is set by the size of the light tunnel from the flash.

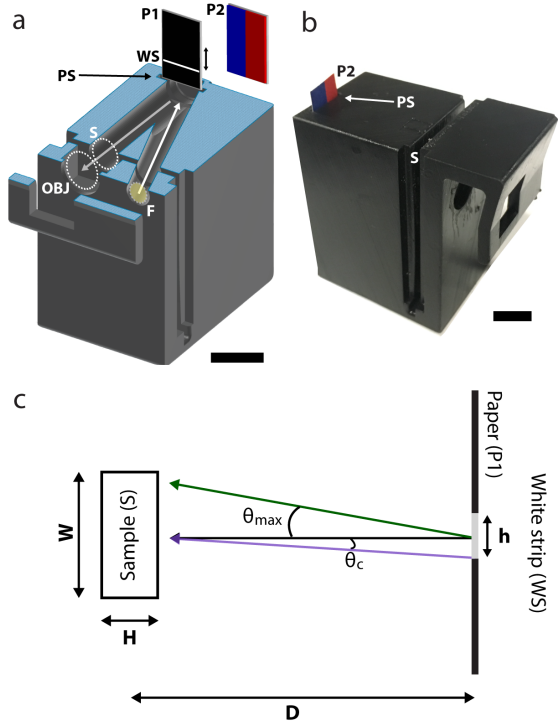

Figure 1. a) A computer-aided design (CAD) visualization of the mobile phone microscope used in this Letter. The CAD model is cut at the blue plane in order to show the internal light tunnel. Light is directed from the flash (F) towards the paper (P1 or P2) which is inserted from the top and can be moved up and down (arrow). P1 and P2 are the two example patterns used in this work. In this schematic, P1 is inserted into the clip's paper slot (PS), and the white strip (WS) diffusely reflects light back towards the sample (S). The blue/red left and right halves of P2 modulate the color of light for left/right projection images. The objective lens (OBJ) approximately collimates light into the mobile phone camera module (module not shown). Scalebar: 10mm. b) 3D printed mobile phone clip. The flash (F), sample (S), paper slot (PS), and paper (P2) locations are indicated. In both (a) and (b), the phone is not shown. Here, the paper Is divided into blue and red halves for stereo imaging. Scalebar: 10mm. c) Ray diagram for when P1 is inserted. Rays incident on the center (black) and edge of the sample (green), subtending an angle $\theta_{max}$. The central ray (black) and the ray connecting the edge of the strip to the center of the sample (purple) subtend the angle $\theta_c$. Physical dimensions: Strip height (h), paper-to-sample distance (D), sample width (W) and height (H). Note the drawing is not to scale for clarity.

When imaging, the mobile phone microscope focus is adjusted so it lies somewhere within the sample depth. A video is then recorded while the paper strip is manually moved up and down, which moves the location of the reflecting strip (Fig. 1a). This in turn changes the illumination angle at the sample. Thus, the image recorded by the mobile phone microscope is a projection of the sample along the angle determined by the location of the diffusely illuminating strip.

Because there is no illumination condenser lens in our design, the angle of a ray at the sample plane resulting from a diffuse source on the paper varies across the field of view (Fig. 1c). This means that, the projection angle will vary across the image, which in turn will affect the parallax and disparity observed at different points in FOV. However, this variation is at most $\theta_{max} = \tan^{-1}[0.143\text{mm}/20\text{mm}] = 0.41°$ between the central and most peripheral regions of the sample (width W=285µm). Parallax causes a disparity, or a translational shift between the same object when illuminated at different angles. This translation is directly proportional to both the height of the sample and the tangent of the illuminating ray. Our sample is H=125µm tall, which yields a disparity shift of $\Delta = (0.143\text{mm}/20\text{mm}) \times 125\text{µm} = 0.89\text{µm}$ at the edge of the FOV for a central projection, for which the disparity shift should be 0µm. This disparity variation is beyond the resolution of our microscope, which has unity optical magnification and 1.22µm pixels[3]. Therefore, the illumination angle can be taken to be uniform over the sample. The illumination distribution at a given point is spread over a small cone of angles $\theta_c = \tan^{-1}[h/2D] = \tan^{-1}[0.5\text{mm}/40\text{mm}] = 0.71°$. This can be adjusted by altering the stripe width on the printed pattern, or redesigning the clip to have a larger sample-to-paper distance D. Along the thin dimension of the illumination strip, the illumination is partly spatially coherent, with a transverse coherence length of $l_c = \frac{1.22 \lambda D}{h} = 13 \mu m$ [6]. Spatial coherence is not exploited in this Letter, however, it may prove possible to do so in the context of Fourier Ptychography, as discussed later on.

Examples of a central and oblique projections are shown in Figs. 2a and b, respectively. The sample is a pair of steps, fabricated in photoresist (IP-S) using a Nanoscribe 3D printer. This photoresist has negligible absorption and a refractive index of ~1.51 at 532nm[9]. Each step has embossed text – "CNBP" on the top step and "RMIT" on the bottom step. The height differential between steps is 50µm and the embossed text is elevated 25µm above each step.

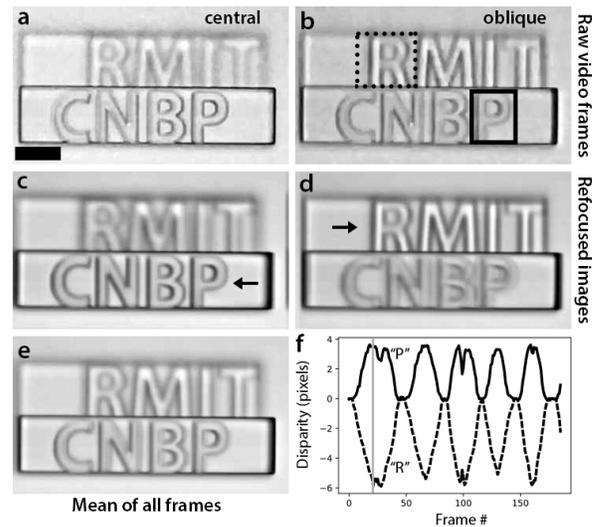

Figure 2. a) Raw video frame when illumination stripe is at the central position. Scale bar: 50µm. b) same as (a) but for an off-axis position of the illumination stripe, yielding an oblique projection of the sample. The dashed box indicates the region used for finding translational shifts between each frame for the "RMIT" text. c) Refocused image at the "CNBP" plane. d) Refocused image at the "RMIT" plane. The arrows in (c) and (d) indicate the refocused text. (e) Mean of all frames in the video. This is equivalent to incoherently averaging the image over all illumination stripe positions. f) The measured disparity translation for the "R" (dashed curve) and "P" (solid curve) letters with respect to the initial central projection. The vertical grey line indicates the frame corresponding to the oblique projection in (b).

As the angle of illumination becomes more oblique, the text features "CNBP" and "RMIT" laterally shift in opposite directions since they are on opposite sides of the focal plane of the mobile phone microscope, yielding a disparity in their apparent positions. This is a manifestation of parallax caused by the angular structure of the illuminating light. This parallax can be used in a manner analogous to light field imaging to refocus the image in post processing. To achieve this, we translate each frame in the video so that either the text "CNBP" or "RMIT" is stationary throughout the entire video. Although the translation vector of each frame is *a priori* unknown due to the manual movement of the paper strip, it can be found by locating the cross-correlation maximum between the first and subsequent frames of the video. Because we are manually translating an illumination strip along one dimension, the translation vector is only along the horizontal direction (though an extension to 2D is trivial). In order to refocus to different planes within the sample, it suffices to multiply the estimated disparity vector for a given plane by a scalar number, since disparity varies linearly with depth. Therefore, only one feature in the dataset is needed to estimate the projection disparity. For Fig. 2, we use the "R" marked by the dashed box in Fig. 2b to estimate the disparity and then rescale to refocus at different planes.

Figures 2c-d show synthetically refocused images refocused at the top and bottom embossed text planes, respectively. A focal stack of 41 refocused planes is included as Supplementary Video 2. Note the reduced noise of Figs. 2c-d compared with Figs. 2a-b due to the synthetically increased condenser aperture size. An overall mean of all frames of the movie, approximating a full condenser aperture image, is shown in Fig. 2e. Here, signal averaging smooths out the image compared with the central projection image in Fig. 2a, however, the depth of field is reduced, causing both the "RMIT" and "CNBP" to be out of focus. The estimated translation along the horizontal direction for the entire movie is shown in Fig. 2a.

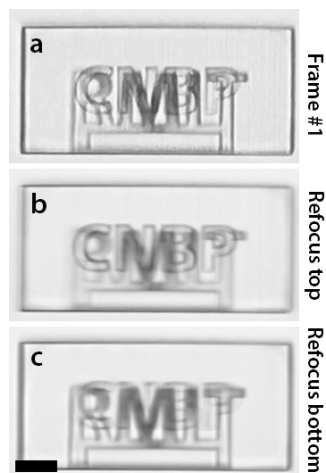

Figure 3. a) Initial raw video frame of overlapping "RMIT" and "CNBP" text sample. The text overlaps in the *xy* plane but is separated by 60μm in height. Text height is 15μm. Sample width is 325μm. Overlapping text is difficult to disambiguate with a single, low illumination numerical aperture (NA) projection. b) Image refocused to the top "CNBP" surface. The bottom "RMIT" text is blurred out of focus. c) Image refocused to the bottom "RMIT" surface. Scale bar: 50μm.

Refocusing is not limited to laterally separated object details. In Figure 3, we demonstrate refocusing on laterally overlapping text on an IP-S photoresist sample. Here, the sample is 75μm tall, with overlapping embossed "CNBP" text on top. The RMIT text is debossed on to the bottom surface. The embossing and debossing height for both texts is 15μm, for a text-to-text separation of 45μm (top of "RMIT" to bottom of "CNBP"). Fig. 3a shows a raw captured video frame with the illuminating strip in a central location. Figs. 3b-c show refocused images of the "CNBP" and "RMIT" text, respectively. Note the blurred "RMIT" text in Fig. 3b, while the "CNBP" text remains in focus, and vice-versa for Fig. 3c.

Figures 2 and 3 demonstrate the utility of dynamic modification of the illumination angle. This approach requires the user to manually move the paper and then calculate the disparity in post-processing. As such the disparity and refocusing information are not available for real time viewing. For single-shot imaging with angle-structured illumination, a different scheme can be employed. Instead of multiplexing angular information in time, we use the color channels in the mobile phone camera to spectrally multiplex different illumination angles. Here we demonstrate this concept applied to stereo microscopy using a static color printed paper pattern. In this instance, we use the P2 paper pattern shown in Figs. 1a-b, where the left half of the paper is printed with blue ink, and the right half with red ink. Upon spatially dependent absorption and diffuse reflection by this piece of paper, the white light illumination from the mobile phone flash is split into blue and red halves travelling in opposite directions towards the sample (Fig. 4a). This is a paper-based version of stereo microscopy previously reported in the optical tweezer literature[10].

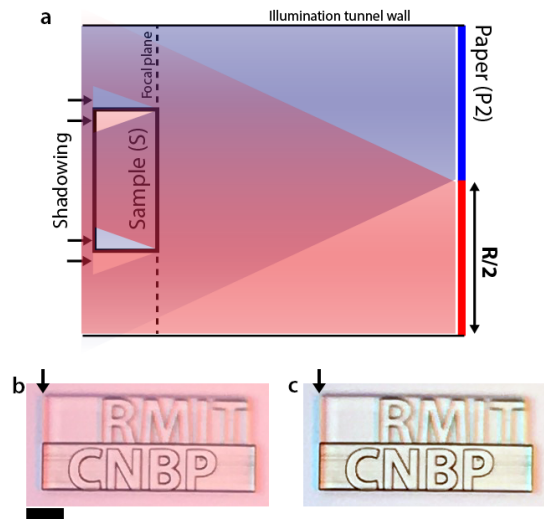

Figure 4. a) Optical schematic of red/blue illumination stereo microscopy. Red/blue shadows at the edges of the sample are indicated with the arrows. This schematic only shows a restricted angular range of the diffuse reflection for clarity. b) Raw stereo image of step object from Fig. 2. The arrow indicates the presence of blue/red shadowing resulting from the spectral-angular structuring of the illumination. A stereo effect is directly observable from this image on the mobile phone

screen when using red/cyan stereo glasses. Scale bar: 50μm. c) The same image as in (b) after correcting white balance for clarity.

With this illumination configuration, the disparity for regions of the sample away from the focal plane is color-coded. Defocused edges appear as blue/red vertical bars, as indicated by the arrows in Figs. 4b-c. The live image displayed by the mobile phone (Fig. 4b) can be directly viewed with red/cyan stereo glasses for a stereo effect. A white-balance corrected image is shown in Fig. 4c for a clearer view of the red/blue color separation.

For high resolution and quantitative stereo microscopy, the paper pattern would ideally be a pair of small horizontally separated red and blue spots on a black background[10]. However, we found that the small amount of reflection from the black ink overwhelmed the diffuse reflection from the red and blue areas (which together have a much smaller area than the black background). As a result, we used the P2 pattern (Fig. 1a) instead which allows the entire illuminated area of the paper to contribute to the color multiplexing effect, albeit with a smaller depth of field and less disparity. Higher order multiplexing using green instead of black ink might circumvent this issue.

We have demonstrated computation refocusing and stereo imaging by structuring illumination with an inkjet-printed piece of paper. Active modulation of the illumination field was achieved by manual translation of the structured paper diffuser and tracked by cross-correlation of recorded video sequences. This ultra-low-tech approach to computational illumination is another route to cost-effective and low-power computational imaging implementations appropriate for field work and educational audiences. Paper patterns are not limited to those demonstrated in this Letter and could even be drawn by hand. For example, multiple paper strips each with a different diameter white circle could be used to vary the illumination numerical aperture (NA) to achieve a desired depth of field. The 3D geometry of the paper reflector could be altered (ie. bent) to further structure the directionality of reflected light.

Potential imaging modalities are not limited to refocusing and stereo. Quantitative differential phase contrast[11,12] and Fourier ptychography[5] imaging data should in principle be able to be captured using our approach. The latter is particularly intriguing as it could in principle increase the achievable resolution in mobile phone microscopes far beyond that supported by cheap, low NA objective lenses typically employed[3,8]. Here, the spatial and spectral coherence of the illumination field would have to be taken into account. With small reflecting areas on the paper strip and a larger paper-to-sample distance, a transverse coherence length of several hundred microns should be possible at the sample plane in a compact format, enabling Fourier Ptychography over a useful FOV [6,7].

**Funding.** Australian Research Council (ARC) (CE140100003).

**Acknowledgment**. We thank the Micro Nano Research Facility at RMIT University for the use of the Nanoscribe 3D printer.